\documentclass[aps,prl,tightenlines,reprint,showpacs,showkeys, 
final,letterpaper]{revtex4-1}

\pdfoutput=1

\usepackage{amsmath,amssymb}
\usepackage[pdftex]{graphicx}

\allowdisplaybreaks

\begin{document}

\title{$J/\psi$ in a hot baryonic plasma}
\author{Purnendu \surname{Chakraborty} }
\author{Subhashis \surname{Chattopadhyay} }
\affiliation{Variable Energy Cyclotron Centre, 1/AF Bidhannagar, Kolkata 
700064, India} 

\begin{abstract}
We calculate the bound state properties of 
$J/\psi$ in a hot and dense QCD plasma using phenomenological potentials 
augmented by inputs from perturbative QCD. The temperature and density 
region of study will be relevant in future heavy ion collision
experiments at FAIR. We find that the effect of baryon density on the 
dissociation of $J/\psi$ is small in this regime. However we  
indicate that if there is a critical end point in the QCD phase diagram then 
strong density fluctuation will dissociate charmonia near hadronization. 
The measurement of $J/\psi$ suppression can therefore signify 
the existence of the critical point unambiguously. 
\end{abstract} 

\maketitle 
The aim of the ongoing relativistic heavy ion collision experiments is to 
create a deconfined phase of strongly interacting matter dubbed as  
quark gluon plasma (QGP). Almost 30 years ago, Matsui and Satz argued that 
the screening of the confining potential at high temperature will 
 lead to the dissolution of heavy quark bound states
in the plasma  and their depleted production 
may be used for a forensic study of the hot and dense medium 
created in such collisions~\cite{Matsui:1986dk}. 
One of the significant results  of SPS heavy ion program was the 
observation of anomalous $J/\psi$ suppression. For $\sqrt{s_{NN}} = 17.3$ GeV 
Pb+Pb and In+In 
collisions, the relative yield of $J/\psi$ was found to be  suppressed  
compared to estimates based on cold nuclear matter (CNM) effects alone 
beyond a centrality threshold~\cite{Alessandro:2004ap,*Arnaldi:2007zz}.
High statistics data for  quarkonia suppression in
Au+Au collisions at $\sqrt{s_{NN}} = 200$ GeV at RHIC~\cite{Adare:2006ns}
and in  
Pb+Pb collisions at $\sqrt{s_{NN}} = 2.76$ TeV at 
LHC~\cite{Pillot:2011zg,*Chatrchyan:2012np,*Chatrchyan:2012lxa} have 
been checked against screening based models and it  
lends strong support for the creation  of a deconfined partonic 
phase~\cite{Grandchamp:2002wp,*Zhao:2011cv,*Gunji:2007uy,*Strickland:2011mw}.

The bulk matter created at RHIC and LHC have low baryon densities, $\mu_B 
\simeq 0$. Upcoming Facility for Antiproton and Ion  Research (FAIR) at GSI 
will collide heavy ions in the energy range $\sqrt{s_{NN}} \sim 6-10$ GeV.
Numerical simulations employing different dynamical models 
show that a medium with high baryon density and relatively low 
temperature is likely to be created at such collision 
energies~\cite{Arsene:2006vf}.   
CBM collaboration at FAIR, in particular,  has a dedicated heavy flavor  
program and is expected to throw light on the possible in-medium 
modifications of open and hidden charm spectra.  At zero baryon density, the  
excited states of charmonium family  - $\chi_c$ and $\psi^\prime$ melt just 
above $T_{pc}^0$ where $T_{pc}^0 \simeq 170$ MeV is the \textit{pseudo-critical}
temperature of QCD in a baryon symmetric medium. 
The ground state may survive upto $\sim 1.5\, T_{pc}^0$ and the maximum 
temperature achievable at FAIR will be below it whereas the chemical potential  
$\mu_q$ could be as high as $\sim 2\,T_{pc}^0$.  
Should we expect to see a density driven melting of $J/\psi$ at low energy 
collisions then? The purport of the present paper is to find an answer.

Heavy quark spectroscopy is well described by  
non-relativistic potential models~\cite{Lucha:1991vn} at zero temperature. 
In statistical QCD, the choice of the potential is debated  and it is 
unclear whether free energy, internal energy or a linear combination thereof   
is the right candidate for it. Nevertheless, potential models have 
been extensively used  at finite temperature to calculate various in-medium 
properties of quarkonia, see~\cite{Mocsy:2013syh,Rapp:2009my} for 
recent reviews. Apart from 
simplicity, the advantage of the potential model is that a
slew of  information can be extracted from it at no 
cost. Such studies complements first principle calculation from  lattice 
gauge theory and seems indispensable now for baryon rich phase of 
QCD where progress in lattice computation is hindered by  
hitherto unsolved sign problem. 

As alluded earlier, our aim here is to understand the in-medium 
modification of charmonia in a hot baryonic plasma which might be produced 
at low energy heavy ion collisions. We scan the region of 
phase diagram where $T/T_{pc}^0 \sim (1 - 1.5)$, and $\mu_q/T_{pc}^0 
\sim \left(1 - 2\right)$  where $T$ and $\mu_q$  are 
equilibrium temperature and chemical potential of the system respectively.
It is assumed that $\mu_u = \mu_d = \mu_q$. How does 
the finite 
baryon chemical potential influence the 
charmonium dissociation?  A large chemical potential implies 
increased screening or weak binding of charmonia in the medium. 
A substantial background temperature is responsible for, apart from 
decrease in binding, rupture of resonances through partonic breakup processes. 
In the extreme case of cold and dense quark matter $\mu_q/T \to \infty$, 
partonic dissociation shuts off and sharp nature of Fermi surfaces may 
lead to nontrivial modification of heavy quark bound 
states~\cite{Kapusta:1988fi}. We relegate this issue for discussion elsewhere.

First quantitative assessment of quarkonia dissociation within potential
model in a hot QCD plasma was made by Karsch, Mehr and 
Satz~\cite{Karsch:1987pv}. For the 
$Q\bar{Q}$ free energy, following choice 
was adopted,     
\begin{equation}
\label{pot1} 
\mathcal{F} = \frac{\sigma}{m_D} \left(1 - e^{-m_D r}\right) - 
\alpha \frac{ e^{-m_D r}}{r}\,.
\end{equation} 
Here $\sigma$ is the string tension and $\alpha = C_F \alpha_s$. 
$C_F = (N_c^2 - 1)/(2 N_c)$ and $\alpha_s$ is the coupling constant of QCD.
$N_c = 3$ is the number of color. $m_D$ is the electric screening mass.   
The long range part of the free energy 
can be realized in Gribov-Zwanziger-Sringl scenario of 
confinement~\cite{Gribov:1977wm,*Zwanziger:1989mf,*Stingl:1985hx} 
involving a  $D = 2$ gluon condensate~\cite{Megias:2007pq,*Riek:2010fk}.

Since the free energy contains an entropy contribution at finite temperature, 
$\mathcal{F} = \mathcal{U} - T S$, it is not  the potential \textit{per se}. 
So it was suggested to use the internal energy instead~\cite{Shuryak:2004tx}. 
Lattice based internal energy were employed in several 
~\cite{Shuryak:2004tx,Alberico:2005xw,*Cabrera:2006wh} investigations. Soon it 
was realized that since the entropy changes rapidly across the transition 
temperature, internal energy computed on lattice provides more binding than 
the vacuum potential. 
In~\cite{Mocsy:2007jz,*Mocsy:2007yj}  the authors constructed a model for 
``maximally binding'' potential by fitting the lattice data 
for free energy at  short and long distances.  
Since a first principle derivation is not possible, we follow here a
simpler approach suggested in~\cite{Dumitru:2009ni}. The internal energy is 
obtained here 
by subtracting entropy (and number density)  contribution at all distances from 
the free energy  in Eq.~\eqref{pot1},
\begin{eqnarray}
\label{pot2}
\mathcal{U} &=& \mathcal{F} - T \frac{\partial \mathcal{F}}{\partial T} 
- \mu_q \frac{\partial \mathcal{F}}{\partial \mu_q} \nonumber \\
&=&\frac{2 \sigma}{m_D} \left(1 - e^{-m_D r}\right)   - 
e^{-m_D r} \left(\sigma r + m_D + \frac{\alpha}{r}\right).   
\end{eqnarray} 
Running of $\alpha_s$ is neglected in arriving at \eqref{pot2}. 
The problem of overshooting the vacuum potential is not eliminated in 
\eqref{pot2} but it is minimal near $T_{pc}^0$ where  
most of the bound states are supposed to melt~\cite{Dumitru:2009ni}.

Recently, Laine and 
collaborators~\cite{Laine:2006ns,*Laine:2007gj,*Burnier:2007qm} 
have shown that the real time static $\bar{Q}Q$ potential has  an 
imaginary part and describes dissociation of quarkonium through scattering 
via exchange of a spacelike gluon (see also \cite{Beraudo:2007ky}). We  
equate the real part of potential to the  
internal energy in Eq.~\eqref{pot2} and augment it by a spin independent 
relativistic correction, the later is  needed for a accurate description of 
charmonium spectrum~\cite{Mocsy:2007jz,Bali:2000gf}.  To wit, real part of 
the potential reads  
$\Re \left\{V\right\} = \mathcal{U} - 0.8\sigma/(m_Q^2 r)$.
 The imaginary part is calculated in hard loop   
approximation~\cite{Laine:2006ns,*Laine:2007gj,*Burnier:2007qm},     
\begin{equation}
\label{pot3}
\Im\left[V\right] = - 2 i \alpha T \int_0^\infty\, \frac{ds
  s}{\left(s^2 + 1\right)^2} 
\left(1 - \frac{\sin{m_D r s}}{{m_D r s}}\right)\,. 
\end{equation}
For the parameters in the potential, we take $\sigma = 0.223$ 
GeV\textsuperscript{2} and $\alpha = 0.385$. 
The electric screening mass is written as, 
\(
m_D^2 = 4 \pi \alpha_s \kappa_1^2 \left(1 + N_f/6\right) T_s^2 
\)
where, 
\begin{equation}
\label{eqforts}
T_s^2 = T^2 \left(1 + \kappa_2 \frac{3N_f}{\pi^2 \left(6 + N_f\right)} 
\frac{\mu_q^2}{T^2} \right)\,.
\end{equation} 
We call $T_s$ an effective screening temperature. It can be thought 
of  
as the equilibrium temperature of a plasma without a net baryon excess  
that produces the same amount of electric screening as the plasma with 
temperature $T$ and chemical potential $\mu_q$. The encapsulation of the 
combined effect of temperature and density in $T_s$ makes
comparison with corresponding result at zero density easier. 
$\kappa_1$ and $\kappa_2$ are parameters to take care of
nonpertubative effects in the transition region.  We take 
$\kappa_1 = 1.4$ as follows from comparing leading order result 
of screening mass with that from a fit to long distance part of 
lattice $Q\bar{Q}$ free energy~\cite{Kaczmarek:2005ui}. 
Determination of $\kappa_2$ is little subtle. On general ground, it is 
expected that  
$\kappa_2 \simeq 1$~\cite{Simonov:2007xc}. 
This is also consistent with lattice result in~\cite{Doring:2005ih} 
for $T \geq 1.5 T_{pc}$. Curiously enough the lattice simulations seem to 
suggest a divergent behavior of $\kappa_2$ and hence a diverging screening 
mass close to $T_{pc}$. Later we shall argue that this divergence in the 
screening mass is a reflection of the proximity to a critical end 
point and discuss the correlated consequences. For the moment being, however, 
we neglect this divergence and  set $\kappa_2 = 1$ in what follows.

\begin{figure}[!htbp]
\includegraphics[width=.7\columnwidth]{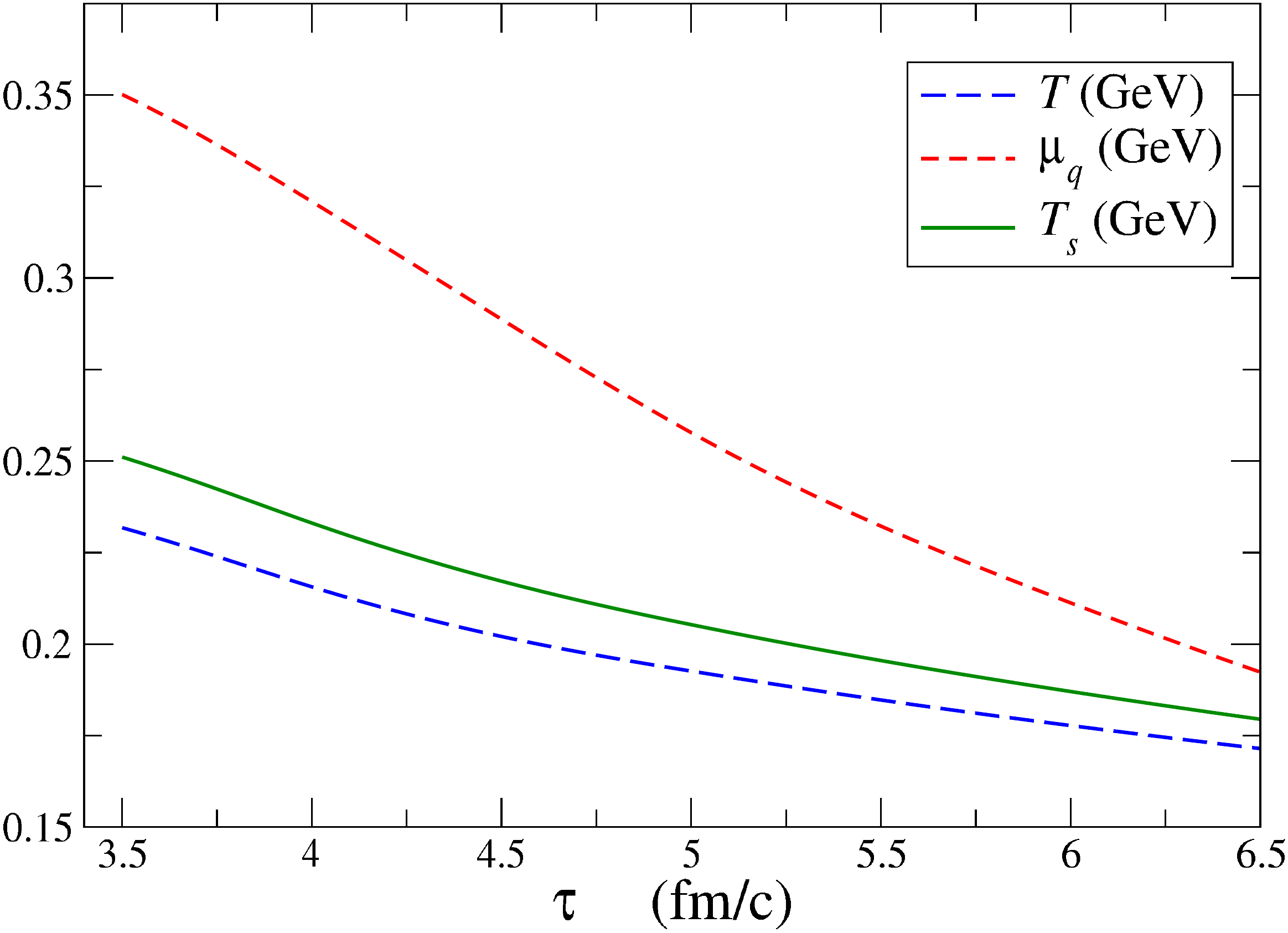}
\caption{\label{fig1}(color online) Evolution of temperature and chemical 
potential in the central hotspot for most central $(b = 0)$  
{Au + Au}  collision at $\sqrt{s_{NN}} = 7.62$ GeV.}
\end{figure}    

Let us now proceed to evaluate bound state properties. 
For definiteness, we 
focus on the central hotspot and consider most central $(b = 0)$ collisions. 
We take the evolution of central baryon density and energy density from 
URQMD simulation at $\sqrt{s_{NN}} = 7.62$ GeV~\cite{Arsene:2006vf}. The 
local temperature and chemical potential 
are then extracted using a ``\textit{fuzzy bag}'' equation of state 
~\cite{Pisarski:2006yk} as parameterized in~\cite{Kapusta:2010ke}. The 
fuzziness here merely represent the $T^2$ correction in 
pressure and could be understood in terms of $D = 2 $ gluon condensate akin to  
the potential in Eq.~\eqref{pot1}~\cite{Megias:2009mp}. We assume 
that the system equilibrates at time $\tau_0 = 3.5$ fm/c which corresponds 
to little more than the passing time of two nuclei 
$\tau = 2 R_A/\sqrt{\gamma^2 - 1}$. The evolution is followed 
until $\tau_f = 6.5$ fm/c when the temperature falls 
below hadronization temperature $T_{pc}^0 = 170$ MeV. 
For brevity, we take here the same hadronization temperature as in zero 
baryon density case.    
The resulting 
evolution of temperature and quark chemical potential is shown 
in Fig.~\ref{fig1}. 
With $N_f = 2$, $T_s \simeq 
T\sqrt{1 + 0.076 \, \mu_q^2/T^2}$ from \eqref{eqforts} and it obviates 
from \eqref{pot2} and \eqref{pot3} 
that the medium effects on the quarkonia  spectroscopy  are   
thus essentially  determined by the background temperature and the density
has little role to play except in extreme conditions. This is transpired in 
Fig.~\ref{fig1} wherein it is shown that $T_s$ remains close to $T$ for the 
entire evolution of the system.

The potential embodied in \eqref{pot2} and \eqref{pot3} 
is now fed into Schr\"{o}dinger equation and complex energy eigenvalue $E = M 
-i\frac{\Gamma}{2}$ is solved for. The binding energy of the 
resonance is obtained as $\epsilon =  
2 m_c + \Re\left\{V\left(r \to \infty\right)\right\} - M$, 
where $m_c = 1.3$ GeV is the charm quark mass. A resonance is effectively 
dissociated when binding energy and decay width come at par 
$\epsilon = \Gamma$. 
 
\begin{figure}[!h]
\centering{
\includegraphics[width=.70\columnwidth]{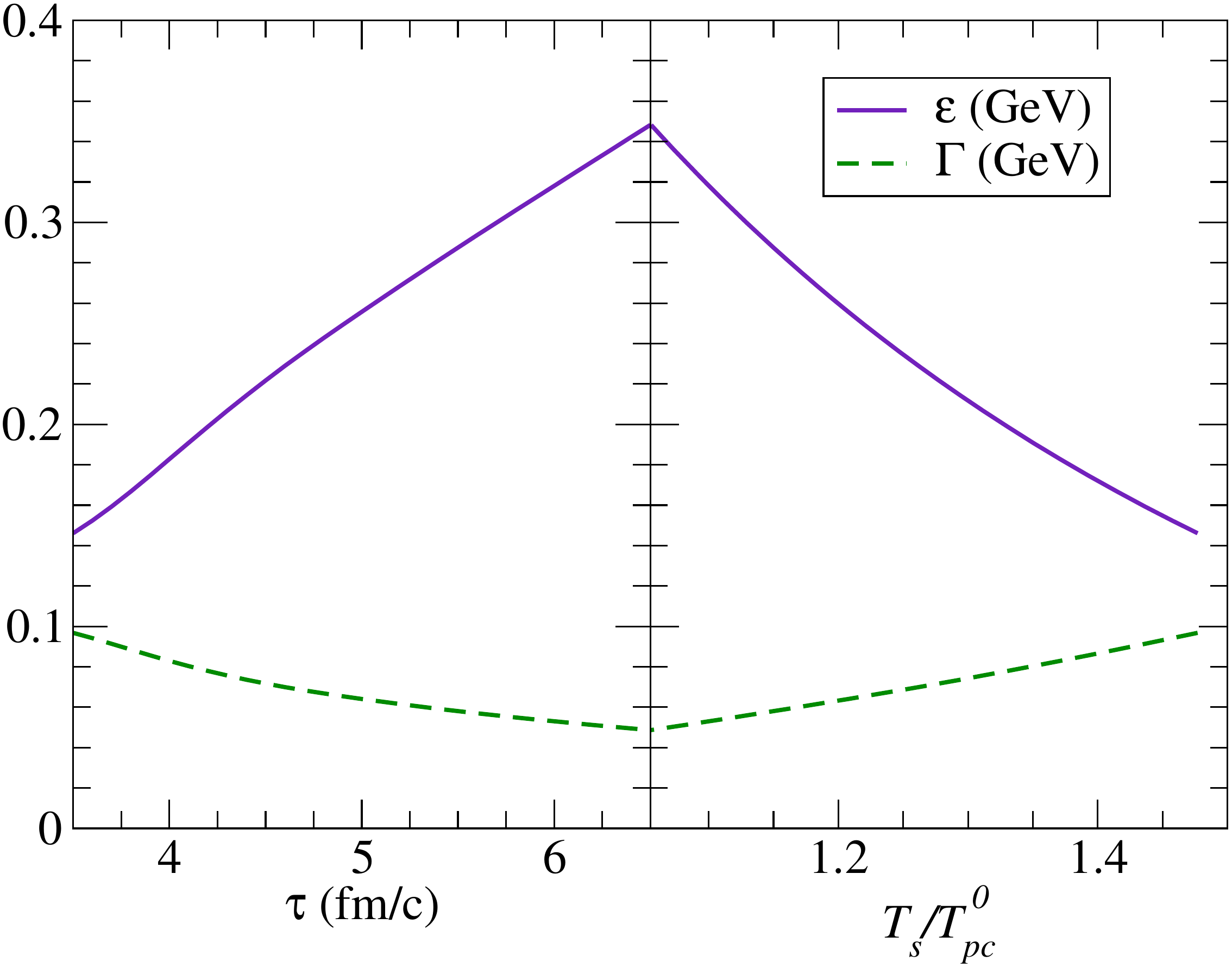}
}
\caption{\label{fig2}(color online) variation of binding energy and decay
width of $J/\psi$ in the central hotspot (left) as a function of time (right) as a 
function $T_s/T_{pc}^0$. For clarity, effect of baryon density is not 
shown separately.} 
\end{figure}

The evolution of binding energy and decay width of $J/\psi$ in the central 
hotspot are shown in 
Fig.~\ref{fig2}. The decay width remains lower than the binding 
energy even at earliest time of evolution when most extreme condition of 
temperature and density are met.  As the system cools and dilutes, 
the screening and decay width become weaker and correlation 
between $Q\bar{Q}$ pair grows until the system hadronizes.  

\begin{figure}[!htbp]
\includegraphics[width=.7\columnwidth]{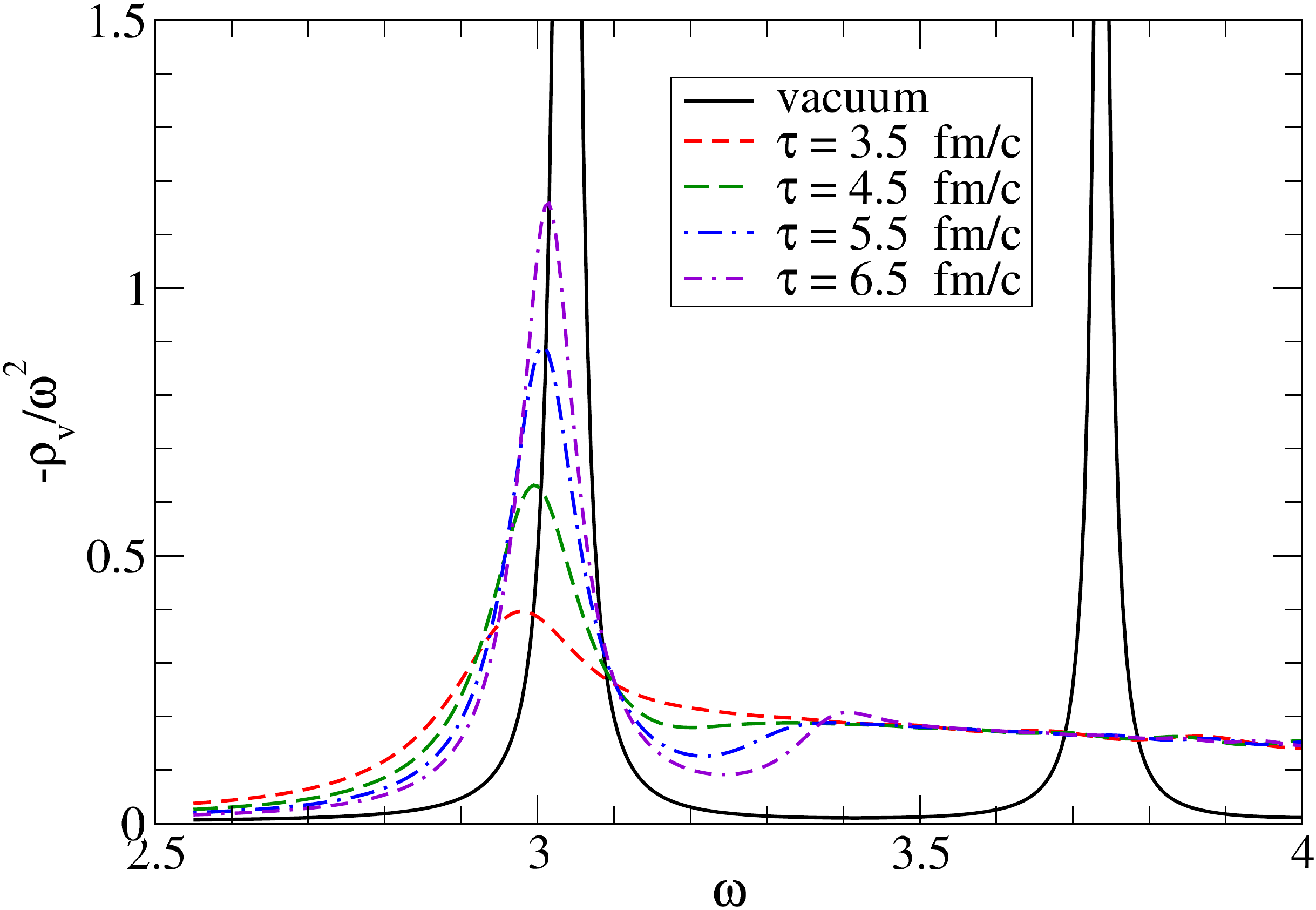}
\caption{\label{fig3}(color online) evolution of $J/\psi$ spectral density}
\end{figure}

We delineate in Fig.~\ref{fig3} the evolution of $J/\psi$ spectral function 
in the central hotspot. 
Close to threshold, the spectral function 
$\rho_v \left(\omega\right)$ is related to the forward correlator, 
\begin{equation}
\rho_v \left(\omega\right) = \lim_{\vec{r},\vec{r}^\prime \to 0}
C_v^{>}\left(\omega, \vec{r}, \vec{r}^\prime\right)\, + \mathcal{O}\left(
e^{-\frac{2m_c}{T}}\right).
\end{equation}
A nice algorithm has been presented in \cite{Burnier:2007qm} for the numerical 
evaluation of the spectral function which we followed here. 

The dissolution of a resonance  
is signaled by the disappearance of the corresponding  peak from the 
spectral function. As seen from the figure, the $J/\psi$ peak is not smeared 
out even at the initial time. The ground state remains strongly 
correlated throughout the evolution. 

The strong correlation between quark-antiquark pair does not imply the 
survival of the resonance in the medium. 
Scattering with the 
particles in the heatbath will destroy this correlation and put the 
quark and antiquark in separate trajectories. 
The pertinent observable here is the 
survival probability, $S = \exp\left(- \int_{\tau_0}^{\tau_f} d\tau\, 
\Gamma \right)$ which measures the fraction of charmonium surviving the 
trek in the medium. Since the density effect on the screening properties of 
the medium is rather small the survival probability is essentially  
determined by the time of exposure of $J/\psi$ to the medium and the 
background temperature. For the energy range covered by FAIR, 
the changes in the local temperature and plasma life time with respect 
to collision energy do not change much so we do not expect appreciable change
in the observed suppression when collision energy is varied. 
At higher collision
energies, competing effect from the regeneration of charmonium in the medium 
becomes important~\cite{Zhou:2013aea}. In fact, the regeneration of charmonium 
is arguably the reason for similar $J/\psi$ suppression at RHIC and SPS.
Combined together, the direct dissociation of
primordial charmonium and the regeneration in the medium is expected to 
result in a rather flat suppression pattern of $J/\psi$ from 
low to moderate collision energy where baryon density 
could have had any effect. If an appreciable deviation of in-medium charmonium 
suppression from the baseline measurement at SPS is observed here then 
it is possibly a hint for a new physics.

What this new physics could be?  Theoretically it has been argued  that 
there is a critical end point (CEP) in the QCD phase diagram where the 
line of first order phase transition  terminates at a second order 
point~\cite{Stephanov:1998dy}. The conjectured critical point 
belongs to the universality class of 3D Ising model. The exact 
coordinate of the CEP on the phase diagram is currently unknown but 
lattice calculations have provided some hazy clue about its 
location~\cite{Fodor:2004nz,*Datta:2012pj}. 
If such a critical point exists it will lead to 
enhanced susceptibilities which can be measured  through 
event-by-event analysis of fluctuation of conserved charges. 
Since the fermionic contribution to the electric screening mass is proportional 
to the quark number susceptibility 
$m_{D,q}^2 \propto \chi_q$~\cite{McLerran:1987pz,*Chakraborty:2001kx},   
an enhancement in susceptibility is also expected to lead to an increased
screening mass. Precisely this behavior is observed in lattice simulation at 
finite chemical potential near $T_{pc}^0$~\cite{Doring:2005ih, Ejiri:2009hq}.
We assume that the critical behavior of  succeptibility 
is also shared by the screening mass  and write 
$(m_D^i)^2 = 2 \pi \alpha_s \chi_q^i/3$, where $m_D^i$ and $\chi_q^i$ are 
irregular part of the electric screening masses and quark number susceptibility 
respectively. 
The divergent part of the quark number number susceptibility is gleaned 
from~\cite{Kapusta:2012zb}, 
\begin{equation}
\label{chi_crit}
\chi_q^i = \frac{9 n_c^2}{\left(\delta + 1\right) P_c} 
\left[\frac{1}{3} \frac{\left(\delta - 1\right)}{\left(2 - \gamma\right)} 
t^\gamma + 5 \delta \left|\eta\right|^{\delta - 1}\right]^{-1}\,. 
\end{equation}
Here, $t = (T - T_c)/T_c$ and $\eta = (n - n_c)/n_c$, $n$ is the 
quark number density and $P$ is the pressure. $T_c$, $P_c$ and $n_c$ are the 
critical values of the respective variables. 
The critical exponents are $\gamma = 1.24$ and 
$\delta = 4.815$.   The regular part of the screening mass has been mentioned in 
\eqref{eqforts}. So we can now see how the presence of a critical 
point inflict upon the survival of $J/\psi$ in the deconfined medium.

\begin{figure}[!htbp]
\includegraphics[width=.70\columnwidth]{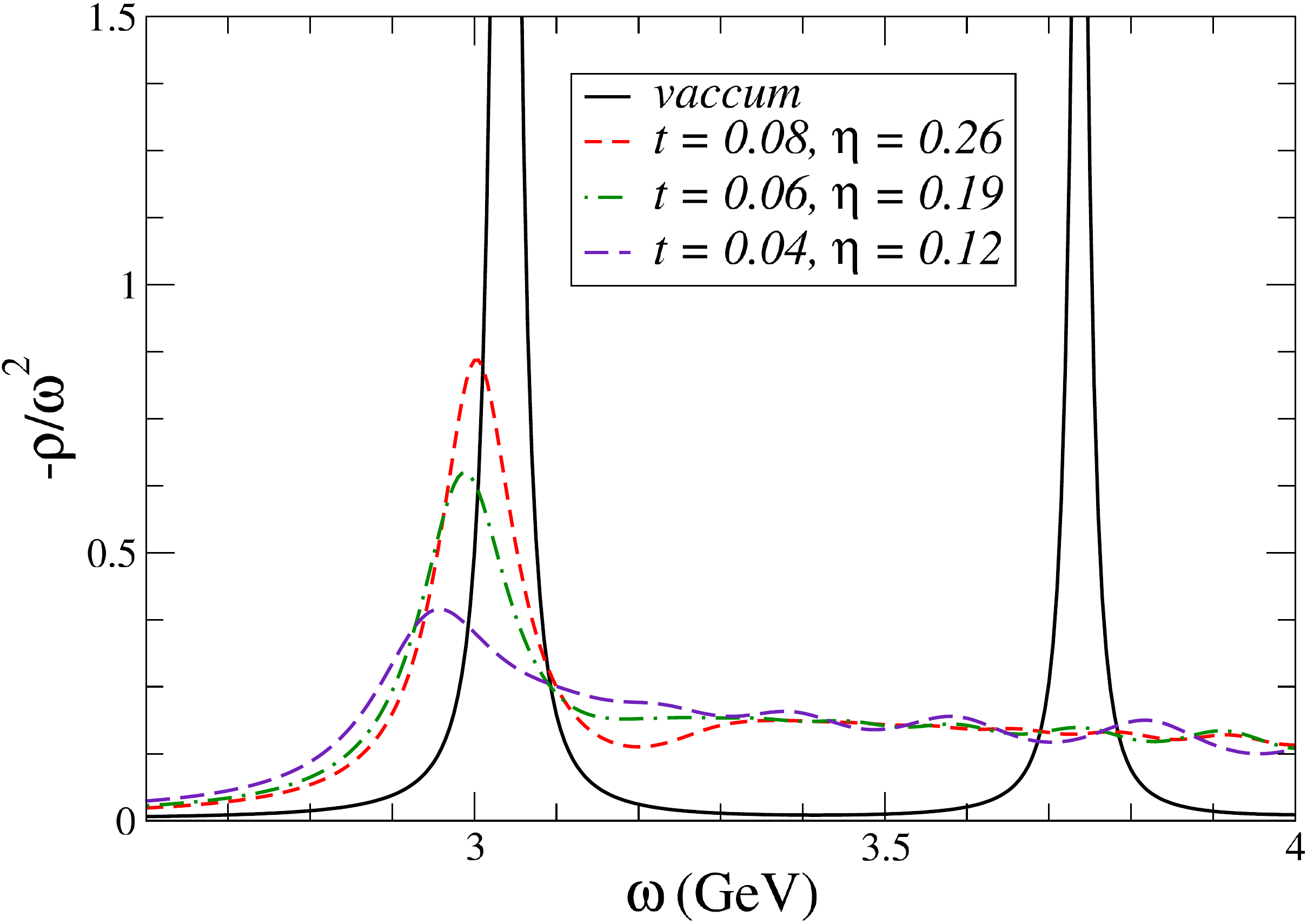}
\caption{\label{fig4}(color online) evolution of $J/\psi$ 
spectral function in presence of a critical point. We chose  
$(T_c, \mu_{c,q)} = (159.8, 138.57)$ MeV. } 
\end{figure}

In  Fig. \ref{fig4} we display the  spectral function of $J/\psi$ near the critical point.
As the critical point is approached, the spectral strength of the $J/\psi$  
is significantly reduced. The loss of 
quark-antiquark correlation  in this case is brought about by the increase of  
screening mass near the critical point. 
This should be contrasted with high $T$ behavior in Fig.~\ref{fig3} 
where disappearance of the spectral peak is caused both by increased screening 
(reduction in strength) as well as increase in decay width 
(smearing of peak). This leads to an interesting picture of charmonium 
survival in a baryonic plasma depending on whether or not the critical point 
is  hit or missed  during the course of evolution. If the evolution of the 
system 
cross the phase boundary away from the critical point then 
the singlet quark-antiquark correlation goes on increasing till
hadronization. On the other hand, the critical point will be hit if the 
evolution of the system proceeds in proximity since CEP acts 
as attractor of hydrodynamical 
trajectories~\cite{Nonaka:2004pg,*Asakawa:2008ti}. As the critical point is 
approached, the $Q\bar{Q}$ correlation goes on dwindling due to 
increased screening and thermal excitation can easily break it off. The 
divergence of the susceptibilities imply that the trajectories in the 
$(T,\mu_q)$ plane 
linger near the  CEP. More the bound state stays close to the CEP more 
likely it is to be broken by scattering in the background medium. It should be 
emphasized that the critical point presents a difficult condition for the 
hidden charm states to be realized close to hadronization.  
By the time the hadronization is complete, the signature of the 
CEP is  imprinted in the near absence of charmonia in the medium and 
it is unlikely that subsequent hadronic evolution will mask it. 
The sudden drop of $J/\psi$ yield at the critical point will therefore provide 
a clean and robust signal for its existence.

Summarizing, we have discussed in detail the bound state properties of 
$J/\psi$ in a hot baryonic medium. This provides the requisite input for 
an all embracing investigation of charmonium production at low collision 
energy. Work along this direction is under progress and will be 
reported elsewhere. Furthermore, we have  argued that if the evolution of 
the medium  proceeds through a critical point then strong density fluctuation 
will remove the charmonium states  from  the spectrum before hadronization. 
This opens up an interesting possibility to locate the critical end point 
through the measurement of charmonium suppression.

\acknowledgements{PC is funded from DAE-SRC award under the scheme No. 2008/21/07-BRNS/2738.}
\bibliography{spectral_refs} 
\end{document}